\documentclass[pra,twocolumn,amsmath,amssymb,superscriptaddress]{revtex4-1}
\usepackage{epsfig,amsmath}
\usepackage{subfigure}
\usepackage{graphicx}
\usepackage{dcolumn}
\usepackage{stmaryrd}
\usepackage{mathrsfs}
\usepackage{pifont}
\usepackage{amsthm}
\usepackage{amssymb}
\usepackage{bm}
\usepackage{latexsym}
\usepackage[colorlinks=true,linkcolor=blue,citecolor=blue]{hyperref}
\usepackage{color}
\usepackage{epstopdf}
\usepackage{booktabs}

\begin{document}

\title{Measurement-induced nuclear spin polarization}

\author{Zhu-yao Jin}
\affiliation{School of Physics, Zhejiang University, Hangzhou 310027, Zhejiang, China}

\author{Jia-shun Yan}
\affiliation{School of Physics, Zhejiang University, Hangzhou 310027, Zhejiang, China}

\author{Jun Jing}
\email{Email address: jingjun@zju.edu.cn}
\affiliation{School of Physics, Zhejiang University, Hangzhou 310027, Zhejiang, China}

\date{\today}

\begin{abstract}
We propose a nuclear-spin-polarization protocol in a general evolution-and-measurement framework. The protocol works in a spin-star configuration, where the central spin is coupled to the surrounding bath (nuclear) spins by flip-flop interaction of equal strength and is subject to a sequence of projective measurements on its ground state. Then a nondeterministic nuclear spin polarization could be implemented by entropy reduction through measurement. The optimized measurement-interval $\tau_{\rm opt}$ is analytically obtained in the near-resonant condition, which is relevant to the nuclear spins' polarization degree of the last-round measurement, the number of nuclear spins, and the coupling strength between the central spin and nuclear spins. Hundreds and even thousands of randomly aligned nuclear spins at the thermal state could be almost fully polarized with an optimized sequence of less than $20$ unequal-time-spacing measurements. In comparison to the conventional approaches, our protocol is not sensitive to the magnetic-field intensity, and it is robust against the extra counterrotating interaction in the near-resonant situation.
\end{abstract}

\maketitle

\section{Introduction}

In scalable solid-state devices for quantum information processing, dynamic nuclear polarization (DNP) is of practical importance for spin-based quantum technology and of fundamental interest to state initialization of nuclear spins~\cite{abragam1961principles,meier2012optical,callaghan1993principles}. For various applications, such as nuclear magnetic resonance, magnetic resonance imaging~\cite{ardenkjaer2003increase,rankin2019recent,denysenkov2017continuous}, discrete-variable quantum computation~\cite{kane1998silicon,schwager2010quantum,kloeffel2013prospect}, and quantum register~\cite{dutt2007quantum}, it is desirable to drive the nuclear spins from an initially thermal state to a fully polarized state.

Various routes toward DNP on transferring polarization from an electron spin to nuclear spins have been actively pursued for a long time. A common theme in all protocols~\cite{abragm1958method,henstra1988nuclear,henstra1988enhanced} that are effective for low electron-spin concentration is the use of a long microwave pulse to match the Larmor frequency of the nuclear spins to the electronic Rabi rotation in the reference frame of the microwave drive, which is well known as a Hartmann-Hahn resonance~\cite{Hartmann1962nuclear}. Under the resonant condition between electron spin and nuclear spin, the hyperfine interaction could play a significant role in polarization transfer. Employing laboratory-frame or rotating-frame level anti-crossings between electron and nuclear spins, there are at least three complimentary mechanisms by which high polarization can be induced in the $^{13}$C nuclear spins in nitrogen-vacancy (NV) center systems: (1) precise control over the external magnetic field in a narrow range~\cite{wang2013sensitive,wang2015strongly,jacques2009dynamic,fischer2013bulk,pagliero2018multispin,
wunderlich2017optically}, (2) magnetic field sweeps~\cite{jacob2019carbon}, and (3) microwave sweeps~\cite{ashok2018orient,ajoy2018enhanced,ajoy2020room,ajoy2021low}. In quantum dots (QD), DNP has been performed in single-dot~\cite{fong2017manipulation}, double-dot~\cite{gullans2010dynamic,schuetz2014nuclear,neder2014theory,Petersen2013large}, and self-assembled-dot~\cite{huang2010theory} systems. The large nuclear spin ensemble ($\sim10^{6}$ spins) could be polarized to an $\approx50\%$ degree under conditions of cryogenic temperature ($T\approx100$ mK) and ultra-strong magnetic field ($B\approx2.9$ T) in a double quantum-dot system~\cite{Petersen2013large}. By tunneling the inter-dot coupling, a nearly $90\%$ polarization has been predicted in theory~\cite{schuetz2014nuclear}.

Under a finite-temperature environment, the external magnetic field breaks the polarization symmetry of nuclear spins in the spatial direction and then transforms the task of DNP to a complete purification of the spins. Inspired by the idea of state purification through repeated projective measurements~\cite{nakazato2003purification}, we consider here DNP in a framework of free evolution and measurement~\cite{Wu2011Nuclear}. In particular, when the ground state of the central spin (electron spin) is closely associated with that of the target spins (nuclear spins), a projective measurement on the ground state of the central spin could force the nuclear spins into their ground state. The strategy is nondeterministic with a finite success probability and has been applied to ground-state cooling in various scenarios~\cite{puebla2020measurement,pyshkin2016ground,yan2021external,yan2022simultaneous}. For a two-spin system under the resonant condition, the protocol can be straightforwardly understood by the effective dynamics of the target spin. Suppose that the central spin $d$ is prepared in its ground state and the target spin $s$ is in an arbitrary state described by a Bloch vector $(r_x, r_y, r_z)$. The probability of the target spin occupying the ground state is $p_g=(1+r_z)/2$. The interaction Hamiltonian reads $H=g(\sigma_s^+\sigma_d^-+\sigma_s^-\sigma_d^+)$, where $g$ is the coupling strength. After a joint evolution with a proper time $\tau$ and measuring the ground state of the central spin, one can find that $p_g \rightarrow p'_g=(1+r_z)/2P>p_g$, where $P=[1+r_z+(1-r_z)\cos^2(g\tau)]/2<1$ is the renormalization constant. By repeating the evolution-and-measurement process with $\cos^2(g\tau)<1$, the population of the target state over the ground state is gradually enhanced. Then after a certain number of rounds, the target spin will approach $(0, 0, 1)$; that is, it is fully polarized or close to it.

In this work, we illustrate how this protocol works with a spin-star model by carrying out the projective measurements on the central (electron) spin. In an ideal situation, the central spin is coupled to the surrounding bath spins with a homogeneous Heisenberg $XY$ interaction. The central spin and bath spins are assumed to be at the ground state $|g\rangle$ and the thermal state, respectively. For a number of bath spins, DNP can be realized through less than a dozen rounds of unequal-time-spacing measurements with optimized measurement intervals. Our protocol is not under the constraint of either a magnetic field in a desired narrow range for NV center systems or a very strong magnetic field for QD systems.

The rest of this work is structured as follows. In Sec.~\ref{HamAndCoolOperater}, we introduce the spin-star model for polarizing a spin bath by repeated measurements. The protocol is generally described by the polarization coefficients, i.e., the occupation reduction factors for the nuclear spins in excited states. In Sec.~\ref{OptimalMeasurementInterval}, we derive an analytical expression for iteratively optimizing the measurement interval in the near-resonant condition and constructing an unequal-time-spacing strategy. In Sec.~\ref{application}, our protocols under both equal-time-spacing and unequal-time-spacing strategies are performed for various sizes of spin bath. Then the optimized unequal-time-spacing strategy is applied to feasible systems, including NV centers and QDs. In Secs.~\ref{Successprobability} and \ref{XYandXYZ}, we discuss the success probability under both strategies and the effects from counterrotating interaction and longitudinal interaction on DNP, respectively. We summarize the whole work in Sec.~\ref{Conclusion}.

\section{Model and Hamiltonian}\label{HamAndCoolOperater}

\begin{figure}[htbp]
\centering
\includegraphics[width=0.95\linewidth]{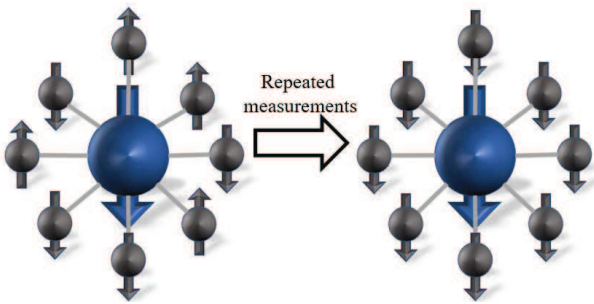}
\caption{Diagram of the spin-star model for our polarization-by-measurement protocol. The central (electron) spin (blue sphere) is homogeneously coupled to the surrounding bath (nuclear) spins (black spheres). The interactions among bath spins are omitted. Initially, the central spin is at the ground state, and the bath spins are at the thermal-equilibrium state. After repeated measurements acting on the central spin, the bath spins could approach a fully polarized state.}\label{model}
\end{figure}

Our polarization protocol is based on the spin-star model shown in Fig.~\ref{model}, which consists of a central spin $1/2$ coupled to $M$ surrounding bath spins-$1/2$ via a Heisenberg $XY$ interaction of equal strength~\cite{jing2007dynamics,Hutton2004mediated,Breuer2004nonmarkovian,Yuan2007nonmarkovian,Radhakrishnan2019Dynamics}. The spins in the bath are identical to and indistinguishable from the central spin. The spin-star configuration is thus a rotationally invariant system which is the direct result of the isotropy of the environment. The full Hamiltonian reads ($\hbar\equiv1$)
\begin{equation}\label{Ham}
H=\frac{\omega_0}{2}\sigma_d^z+\frac{\omega_1}{2}\sum_{j=1}^M\sigma_j^z
+g\sum_{j=1}^M\left(\sigma_d^x\sigma_j^x+\sigma_d^y\sigma_j^y\right),
\end{equation}
where $\omega_0$ and $\omega_1$ are the frequencies of the central and bath spins, respectively, $\sigma^{x,y,z}$ are Pauli operators, and $g$ is the homogeneous coupling strength between them. In the mean-field approach or the ``box-model''  for electron-nuclei interaction~\cite{Chesi2015theory}, $g$ is usually introduced as an average over the hyperfine constants between the central spin and individual nuclear spin. The energy spacing for the electron spin is much larger than the nuclear spins as well as the coupling strength by several orders. It is thus a reasonable idealization for describing solid-state systems~\cite{Urbaszek2013nuclear}. In the rotating frame with respect to $H'_0=\omega_1/2(\sigma_d^z+\sum_{j=1}^M\sigma_j^z)$, the Hamiltonian can be written as
\begin{equation}\label{IntHam}
\begin{aligned}
H'&=e^{iH'_0t}H{e^{-iH'_0t}}-H'_0\\
&=\frac{\Delta}{2}\sigma_d^z+2g\sum_{j=1}^M\left(\sigma_d^+\sigma_j^-+\sigma_d^-\sigma_j^+\right),
\end{aligned}
\end{equation}
where $\Delta=\omega_0-\omega_1$ is the detuning between the central spin and bath spins and $\sigma^+=|e\rangle\langle g|$ and $\sigma^-=|g\rangle\langle e|$ are the transition operators. Using the collective angular momentum operators $J_{\pm}\equiv\sum_{j=1}^M\sigma_j^{\pm}$~\cite{wang2013non,hamdouni2006exact}, we have
\begin{equation}\label{BigspinHam}
H'=\frac{\Delta}{2}\sigma_d^z+2g\left(J_+\sigma_d^-+J_-\sigma_d^+\right).
\end{equation}
To obtain a compact analytical expression that can be used to predict the efficiency of our polarization protocol, we here take a further approximation to ignore the degeneracy of bath spins with the same excitation number by virtue of their identity. It is equivalent to consider only the subspace spanned by the states with maximum total angular momentum $J=M/2$, similar to the Dicke model~\cite{Dicke,Hutton2004mediated}. Then the eigenstates of the spin bath can be denoted by the eigenbasis $\{|m\rangle\}$ of $J_z\equiv\sum_{j=1}^M\sigma_j^z/2$, where $m$ runs from $0$ to $M$, indicating the excited number of bath spins~\cite{Coish2004Hyperfine}. $|m=0\rangle$ implies that $M$ bath spins are all in the ground state, and $|m=M\rangle$ implies that they are all in the excited state. Both of them are fully polarized, but their symmetry is broken by a finite magnetic field. The collective angular momentum operators~\cite{hamdouni2006exact} satisfy
\begin{equation}\label{Jformula}
\begin{aligned}
J_z|m\rangle&=\left(m-\frac{M}{2}\right)|m\rangle,\\
J_+|m\rangle&=\sqrt{(M-m)(m+1)}|m+1\rangle,\\
J_-|m\rangle&=\sqrt{(M-m+1)m}|m-1\rangle.
\end{aligned}
\end{equation}

The central spin and the bath spins are supposed to be initially separable and respectively in the ground state and the thermal state with a finite temperature $T$, i.e., $\rho(0)=|g\rangle\langle g|\otimes\rho_s(0)$. Using Eq.~(\ref{Jformula}), the initial state of the bath spins can be written as
\begin{equation}\label{initialbathState}
\rho_s(0)=\sum_{m=0}^{M}p_m|m\rangle\langle m|, \quad
p_m=\frac{1}{Z}e^{-\beta\omega_1(m-M/2)},
\end{equation}
where $Z\equiv{\rm Tr}[\exp(-\beta\omega_1J_z)]$ is the partition function and $\beta=1/(k_BT)$ is the inverse temperature of the spin bath, with $k_B$ being the Boltzmann constant.

In the framework of free evolution and measurement, our DNP protocol is performed through rounds of joint free evolution $U(\tau)=\exp(-iH'\tau)$ under the interaction Hamiltonian in Eq.~(\ref{BigspinHam}) with a time spacing $\tau$ and instantaneous projective measurement $\mathcal{M}\equiv|g\rangle\langle g|$ acting on the ground state of the central spin. Note $|g\rangle|m=0\rangle$ is the global ground state for the Hamiltonian in either Eq.~(\ref{IntHam}), which covers the whole space with varing total angular momentum $J$, or Eq.~(\ref{BigspinHam}), which involves only the subspace with $J=M/2$. All the excited states are distributed in the subspaces ordered by nonzero excitation numbers. The repeated projections over $|g\rangle$ of the central spin therefore dramatically change the populations of the bath spins, by discarding their distributions in the manifolds except $|0\rangle^{\otimes M}$ or $|m=0\rangle$, i.e., individual spins in Eq.~(\ref{IntHam}) or the larger spin in Eq.~(\ref{BigspinHam}). If the outcome of the measurement is that the central spin is not in $|g\rangle$, then the system sample is abandoned, and the protocol restarts from the beginning. This strategy is equivalent to reducing the entropy of the whole system conditionally by quantum measurement.

Under the equal-time-spacing strategy with $N$ rounds of evolution-and-measurement, the bath state turns out to be
\begin{equation}\label{rhosNtau}
\rho_s(N\tau)=\frac{V(\tau)^N\rho_s(0)V^{\dagger}(\tau)^N}{P(N)},
\end{equation}
where $V(\tau)=\langle g|U(\tau)|g\rangle$ constitutes a nonunitary time-evolution operator for the bath spins and $P(N)={\rm Tr}[V(\tau)^N\rho_s(0)V^{\dagger}(\tau)^N]$ is the success probability of finding the central spin in its ground state $|g\rangle$ at time $t=N\tau$. In terms of the eigenbasis $\{|m\rangle\}$, we have
\begin{equation}\label{V}
V(\tau)=\sum_{m=0}^M\alpha_m(\tau)|m\rangle\langle m|,
\end{equation}
where $\alpha_m(\tau)$ is the polarization coefficient describing the population-reduction ratio on the state $|m\rangle$,
\begin{equation}\label{CoolingCoeff}
\begin{aligned}
\alpha_m(\tau)&=\cos\left(\Omega_m\tau\right)
+i\frac{\Delta\sin\left(\Omega_m\tau\right)}{2\Omega_m}, \\
\Omega_m&\equiv\sqrt{\Delta^2/4+4g^2m(M-m+1)}.
\end{aligned}
\end{equation}
Using Eqs.~(\ref{initialbathState}), (\ref{rhosNtau}), and (\ref{V}), we have
\begin{equation}\label{rhoNtauwithalpha}
\begin{aligned}
\rho_s(N\tau)&=\frac{\sum_{m=0}^M|\alpha_m(\tau)|^{2N}p_m|m\rangle\langle m|}{P(N)},\\
P(N)&=\sum_{m=0}^M|\alpha_m(\tau)|^{2N}p_m.
\end{aligned}
\end{equation}

\begin{figure}[htbp]
\centering
\includegraphics[width=0.95\linewidth]{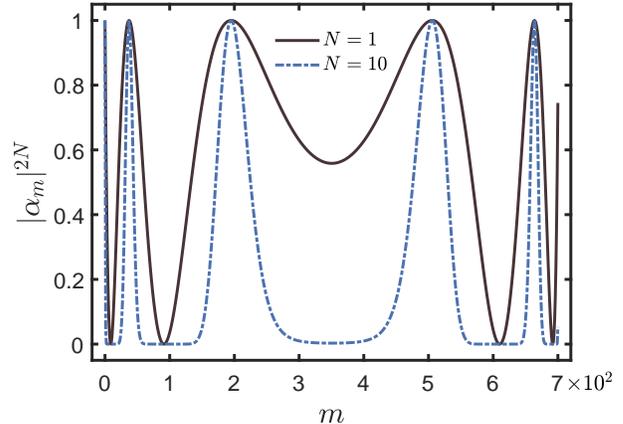}
\caption{Polarization coefficient $|\alpha_m(\tau)|^{2N}$ as a function of the eigenbasis index (the magnetic quantum number) $m$ for $J_z$ by a single measurement (black solid line) and $10$ equal-time-spacing measurements (blue dot-dashed line) on the ground state of the central spin coupled to $M=700$ bath spins. The detuning between the central spin and bath spins is $\Delta/\omega_0=0.1$, the coupling strength is $g/\omega_0=0.1$, and the measurement interval is $\omega_0\tau=0.03$.}\label{coefficient}
\end{figure}

Our polarization-by-measurement protocol is self-content because $|\alpha_m(\tau)|^2\leq1$, where the equivalence is achieved when $m=0$ or $\Omega_m\tau=k\pi$. The populations over the other states are gradually reduced by $p_m\rightarrow|\alpha_m(\tau)|^{2N}p_m$. The reduction rate is clearly determined by $\tau$ due to Eq.~(\ref{CoolingCoeff}). Although the probability of the bath spins in the fully polarized state $|m=0\rangle$ could be significantly increased by repeated measurements, the polarization coefficients shown in Fig.~\ref{coefficient} indicate that under the strategy of equal-time-spacing measurements, there will always be several excited states that are protected. A thermal state with a finite temperature will thus be reduced to a classical mixture of a fully polarized state $|m=0\rangle$ and those states satisfying $\Omega_m\tau=k\pi$. Note $|0\rangle^{\otimes M}$ or $|m=0\rangle$ is both a fully polarized state and a ground state of the spin-bath. Our protocol always holds even when considering all the other subspaces of the excited states as long as $p_0\neq0$, $|\alpha_{m=0}(\tau)|^2=1$, and $|\alpha_{m\neq0}(\tau)|^2<1$, which must be true for an initial thermal state. In the example provided in the Appendix, one can see that we require more numbers of measurements to attain the same degree of polarization if we work in the whole space. We are then motivated to find an optimized measurement interval $\tau_{\rm opt}$ and employ an unequal-time-spacing strategy to improve the performance of our polarization protocol.

\section{Optimized measurement-interval and unequal-time-spacing strategy}\label{OptimalMeasurementInterval}

To see more clearly the effect of the measurement interval $\tau$ on DNP, we first define a polarization degree of bath spins as
\begin{equation}\label{defpolarization}
\mathcal{P}(t)=\left|\frac{{\rm Tr}[J_z\rho_s(t)]}{M/2}\right|=\frac{\sum_{m=0}^Mp_m(t)(M/2-m)}{M/2},
\end{equation}
where $p_m(t)$ is the current population over $|m\rangle$. $\mathcal{P}(t)$ ranges from $0$ (the most mixed state) to $1$ (the fully polarized state) and is consistent with the previous definition~\cite{Deng2006analytical,Morley2007efficient}, i.e., $\mathcal{P}=(M_{\uparrow}-M_{\downarrow})/M$, where $M_{\uparrow}$ ($M_{\downarrow}$) is the number of nuclear spins in the up (down) states. Using Eq.~(\ref{rhoNtauwithalpha}), we have
\begin{equation}\label{polarization}
\mathcal{P}(N)\equiv\mathcal{P}(t=N\tau)=\frac{\sum_{m=0}^M(M/2-m)|\alpha_m|^{2N}p_m}{M/2\sum_{m=0}^M|\alpha_m|^{2N}p_m}.
\end{equation}

\begin{figure}[htbp]
\centering
\includegraphics[width=0.95\linewidth]{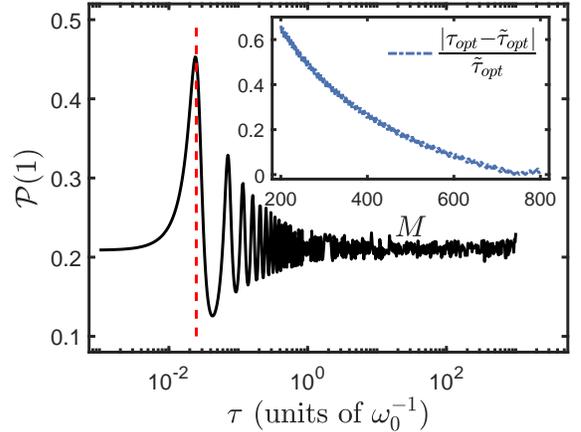}
\caption{Polarization degree of $M=700$ bath spins after one measurement as a function of the measurement interval $\tau$. The black solid curve is numerically obtained using Eq.~(\ref{polarization}). The vertical red dashed line is the analytical result $\tau_{\rm opt}$ in Eq.~(\ref{opttauP}). Inset: Relative error of the optimized analytical interval $\tau_{\rm opt}$ compared with the numerical result $\tilde{\tau}_{\rm opt}$ as a function of the bath-spin number $M$. $\Delta/\omega_0=0.1$, $g/\omega_0=0.1$, $T=0.5$ K and $\omega_0=100$ MHz.}\label{tau}
\end{figure}

A quantitative observation about the effect of the measurement interval on $\mathcal{P}$ is presented in Fig.~\ref{tau} for $N=1$ and $M=700$. The polarization degree is not a monotonic function of $\tau$. It increases rapidly with increasing $\tau$ to the maximum value until an optimized measurement interval $\tau_{\rm opt}$ and then decreases abruptly to a lower value than that determined by the initial temperature. Afterwards, it fluctuates with a decreasing magnitude and asymptotically approaches the initial polarization. Thus, an inappropriate choice of the measurement interval $\tau$ yields either inefficient polarization or even depolarization. To locate the optimized $\tau$ for the highest $\mathcal{P}(1)$ in proximity to the dramatic-change point along the curve, it is instructive to find a local minimum of the denominator in Eq.~(\ref{polarization}) with $N=1$, which is a summation over $|\alpha_m|^2$ with the weight $p_m$. The occupation probability $p_m$ given by Eq.~(\ref{initialbathState}) declines monotonically with increasing $m$, and around $m=0$ and $\tau=0$, the polarization coefficient can be approximated by
\begin{equation}\label{alpham2}
|\alpha_m(\tau)|^2=1-\Omega_m'^2\tau^2+\left(\Omega_m'^2+\frac{\Delta^2}{4}\right)
\Omega_m'^2\frac{\tau^4}{3}+O(\tau^6)
\end{equation}
with $\Omega_m'\equiv2g\sqrt{m(M-m+1)}=\sqrt{\Omega_m^2-\Delta^2/4}$. Thus the polarization degree is rewritten as
\begin{equation}\label{approxpolarization}
\begin{aligned}
\mathcal{P}(1)&=\frac{\sum_{m=0}^M(M/2-m)x^m|\alpha_m|^2}{M/2\sum_{m=0}^Mx^m|\alpha_m|^2}\\
&\approx\frac{\sum_{m=0}^M(M/2-m)x^m|\alpha_m|^2}{M/2\sum_{m=0}^Mx^m(1-\Omega_m'^2\tau^2)}\\
&\approx\frac{\sum_{m=0}^M(M/2-m)x^m|\alpha_m|^2/(M/2)}{\sum_{m=0}^{\infty}x^m(1-\Omega_m'^2\tau^2)}\\
&=\frac{\sum_{m=0}^M(M/2-m)x^m|\alpha_m|^2/(M/2)}{\sum_{m=0}^{\infty}x^m+4g^2\tau^2\sum_{m=0}^{\infty}[m^2-(M+1)m]x^m}\\
\end{aligned}
\end{equation}
with $x\equiv\exp(-\beta\omega_1)$. An approximate ``singularity'' for Eq.~(\ref{approxpolarization}) emerges as
\begin{equation}\label{opttauP}
\begin{aligned}
\tau_{\rm opt}&=\sqrt{\frac{\sum_{m=0}^{\infty}x^m}{4g^2\sum_{m=0}^{\infty}[(M+1)mx^m-m^2x^m]}}\\
&=\sqrt{\frac{1}{4g^2[(M+1)x/(1-x)-(1+x)x/(1-x)^2]}}\\
&=\frac{1}{gM\sqrt{2(1-\mathcal{P}_{\rm th})\mathcal{P}_{\rm th}}},
\end{aligned}
\end{equation}
where we have used the geometric series
\begin{equation}
\sum_{m=0}^{\infty}m^2x^m=\frac{(1+x)x}{(1-x)^3}, \quad \sum_{m=0}^{\infty}mx^m=\frac{x}{(1-x)^2},
\end{equation}
and $\mathcal{P}_{\rm th}$ is the initial polarization degree,
\begin{equation}\label{Pth}
\begin{aligned}
\mathcal{P}_{\rm th}&=\frac{\sum_{m=0}^M(M/2-m)p_m}{M/2\sum_{m=0}^Mp_m}
\approx\frac{\sum_{m=0}^{\infty}(M/2-m)x^m}{M/2\sum_{m=0}^{\infty}x^m} \\ &=1-\frac{2x}{M(1-x)}.
\end{aligned}
\end{equation}

Note that the upper bound $M$ for certain summations in Eqs.~(\ref{approxpolarization}), (\ref{opttauP}), and (\ref{Pth}) has been approximated by infinity to attain a compact analytical expression such that the singularity from a vanishing denominator in $\mathcal{P}(1)$ does not really exist and $\tau_{\rm opt}$ is then an estimation used to locate a maximum $\mathcal{P}(1)$. The second-order perturbative optimized measurement interval $\tau_{\rm opt}$ is irrelevant to the detuning $\Delta$ between the central spin and the bath spins due to Eq.~(\ref{alpham2}), so that Eq.~(\ref{opttauP}) applies to both resonant and near-resonant situations. $\tau_{\rm opt}$ is marked by the vertical red dashed line in Fig.~\ref{tau}, which matches perfectly the point for catching a peak value for $\mathcal{P}$. The inset in Fig.~\ref{tau} describes the relative error between analytical and numerical results for the optimized interval, $|\tau_{\rm opt}-\tilde{\tau}_{\rm opt}|/\tilde{\tau}_{\rm opt}$, as a function of the bath-spin number $M$. The error magnitude decreases roughly with increasing $M$. When $M\geq 580$, it becomes less than $10\%$. When $M=700$, it is about $3.7\%$, consistent with the result in the main plot in Fig.~\ref{tau}.

Both population distributions $p_m$ over the eigenstates $\{|m\rangle\}$ and the polarization degree $\mathcal{P}$ of the spin bath would be modified after the first round of evolution and measurement with an interval $\tau_{\rm opt}$ determined by $\mathcal{P}_{\rm th}$ in Eq.~(\ref{Pth}). The optimized measurement-interval expression in Eq.~(\ref{opttauP}) for $\tau_{\rm opt}$ is then no longer appropriate. An unequal-time-spacing strategy therefore emerges from iteratively updating $\mathcal{P}_{\rm th}$ with the polarization degree of the last round. Consequently, Eq.~(\ref{opttauP}) can be reinterpreted as
\begin{equation}\label{unequaltau}
\tau_{\rm opt}(t)=\frac{1}{gM\sqrt{2[1-\mathcal{P}(t)]\mathcal{P}(t)}},
\end{equation}
where $\mathcal{P}(t)$ represents the current polarization degree of bath spins. Given $\tau_{\rm opt}(t)$ of the last round, the density matrix of the spin bath can be obtained using Eq.~(\ref{rhoNtauwithalpha}) and subsequently the current polarization degree is calculated using Eq.~(\ref{defpolarization}) without a realistic measurement. Rather than a constant $\tau_{\rm opt}$, Eq.~(\ref{unequaltau}) gives rise to a time-dependent sequence: $\{\tau_{\rm opt}(t_1), \tau_{\rm opt}(t_2), \cdots, \tau_{\rm opt}(t_N)\}$, with $t_{i>1}=\sum_{j=1}^{j=i-1}\tau_{\rm opt}(t_j)$ and $\tau_{\rm opt}(t_1)=\tau_{\rm opt}$. For Eq.~(\ref{unequaltau}), when $\mathcal{P}(t)$ approaches unit during the DNP process, $\tau_{\rm opt}(t)$ becomes even larger, meaning that a further polarization becomes more difficult.

Under the unequal-time-spacing protocol, the state of the spin-bath in Eq.~(\ref{rhoNtauwithalpha}) is transformed to
\begin{equation}\label{unequalrho}
\rho_s\left[\sum_{i=1}^{N}\tau_{\rm opt}(t_i)\right]=\frac{\sum_{m=0}^M\prod_{i=1}^{N}|\alpha_m[\tau_{\rm opt}(t_i)]|^2p_m|m\rangle\langle m|}{P(N)}
\end{equation}
after $N$ measurements, and the success probability becomes
\begin{equation}\label{unequalPg}
P(N)=\sum_{m=0}^M\prod_{i=1}^N\left|\alpha_m\left[\tau_{\rm opt}(t_i)\right]\right|^2p_m.
\end{equation}
Now one can find that the time variable for the polarization coefficient $\alpha_{m}$ becomes time dependent and then all the excited states are no longer protected when $N>1$. The unequal-time-spacing protocol is thus more efficient than its equal-time-spacing counterpart in polarization.

\section{Polarization performance}\label{application}

\subsection{Polarization performance under the near-resonant condition}

In this section, we demonstrate the polarization performance of the bath spins in the NV-center system~\cite{Sangtawesin2016} with the equal-time-spacing and unequal-time-spacing polarization-by-measurement strategies under the near-resonant condition. Accordingly, the optimized measurement interval $\tau_{\rm opt}$ is then given by Eq.~(\ref{opttauP}) or (\ref{unequaltau}). In numerical evaluations, the eigenfrequency of the central (electron) spin is chosen to be $\omega_0=120$ MHz. The detuning between the central spin and bath spins and their coupling strength are fixed to $\Delta/\omega_0=0.1$ and $g/\omega_0=0.03$, respectively. And the spin bath is initialized with a temperature $T=0.5$ K.

\begin{figure}[htbp]
\centering
\includegraphics[width=0.95\linewidth]{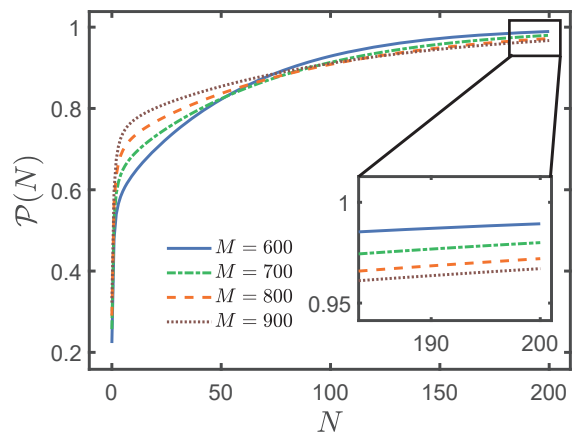}
\caption{Polarization degree of nuclear spins $\mathcal{P}(N)$ as a function of the measurement number $N$ under the equal-time-spacing strategy with varying size of the nuclear spin-bath. The blue solid line, the green dot-dashed line, the orange dashed line, and the brown dotted line represent $M=600, 700, 800$, and $900$, respectively. The other parameters are set as $\Delta/\omega_0=0.1$, $g/\omega_0=0.03$, and $T=0.5$ K.}\label{performance}
\end{figure}

Figure~\ref{performance} demonstrates the performance of the equal-time-spacing strategy for a spin bath with varying size. According to Eqs.~(\ref{opttauP}) and (\ref{Pth}), a larger size of spin bath yields a smaller $\tau_{\rm opt}$ and a higher initial polarization degree $\mathcal{P}_{\rm th}$. In particular, for $M=600$, $\mathcal{P}_{\rm th}=0.223$; for $M=700$, $\mathcal{P}_{\rm th}=0.257$; for $M=800$, $\mathcal{P}_{\rm th}=0.290$; and for $M=900$, $\mathcal{P}_{\rm th}=0.322$. In the first few dozens of rounds of measurements, the polarization rate of a larger size of spin bath is higher than that of a smaller size. And the former becomes lower than the latter as more measurements are carried out. At around $N=75$, the four curves cross each other. With even more measurements, a smaller $M$ yields a slightly bigger asymptotic value of $\mathcal{P}(N)$. When $N=200$, the inset of Fig.~\ref{performance} shows that for $M=600$, $\mathcal{P}=0.989$; for $M=700$, $\mathcal{P}=0.980$; for $M=800$, $\mathcal{P}=0.972$; and for $M=900$, $\mathcal{P}=0.967$. On the whole, the polarization degrees $\mathcal{P}(N)$ can be enhanced from their initial values to nearly unit by a sufficiently large number of rounds of evolution and measurement. The decreasing polarization rates under the equal-time-spacing strategy indicate explicitly that the optimized measurement interval determined by the initial thermal-state polarization degree $\mathcal{P}_{\rm th}$ becomes even more inefficient for the subsequent rounds of measurement, as can be predicted by Eq.~(\ref{opttauP}).

\begin{figure}[htbp]
\centering
\includegraphics[width=0.95\linewidth]{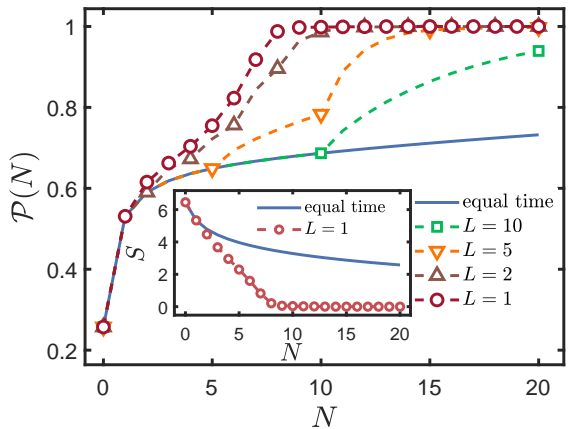}
\caption{Polarization degree of $M=700$ nuclear spins as a function of measurement number $N$ for various polarization strategies. The blue solid line represents the equal-time-spacing strategy. The green dashed line with squares, the orange dashed line with inverted triangles, the brown dashed line with triangles, and the red dashed line with circles represent the strategies in which the measurement interval is updated every $L=10, 5, 2, 1$ rounds of evolution and measurement, respectively. Inset: The von Neumann entropy $S$ of the spin bath as a function of measurement number $N$. The other parameters are the same as in Fig.~\ref{performance}.}\label{UnequalTime}
\end{figure}

The equal-time-spacing strategy is optimized in only the first round of the sequence. To enhance the polarization performance by accurately locating every peak value of the polarization degree under one measurement, one has to iterate the optimized measurement interval according to Eq.~(\ref{unequaltau}). In Fig.~\ref{UnequalTime}, we present the dynamics of the polarization degree for $M=700$ nuclear spins under the equal-time-spacing strategy and four unequal-time-spacing strategies with various iterative rates $L$. For example, $L=5$ means that $\tau_{\rm opt}(t)$ is updated every five rounds of evolution and measurement. Accordingly, the equal-time-spacing strategy means $L\rightarrow\infty$. For $L=1$, we have $\tau_{\rm opt}(t_i)<\tau_{\rm opt}(t_{i+1})$ in the realistic sense of the unequal-time-spacing strategy. It is observed that more updating of the optimized measurement interval gives rise to better polarization performance. In particular, one has to run the strategy with $L=5$ for $N=15$ rounds or run that with $L=1$ for only $N=8$ rounds to achieve $\mathcal{P}>0.99$. In comparison to the strategy of equal time spacing in Fig.~\ref{performance}, the number of measurements is reduced by one order under that of unequal time spacing, indicating a dramatic advantage in experimental overhead.

The effect of polarization from measurements can be understood by the dynamics of the von Neumann entropy of the bath spins. It is evaluated by
\begin{equation}\label{entropy}
S\left[\rho_s(t)\right]=-\sum_{m=0}^{M}p_m(t)\ln p_m(t),
\end{equation}
where the spin-bath density matrix $\rho_s$ is given by Eq.~(\ref{rhoNtauwithalpha}) and by Eq.~(\ref{unequalrho}) under the equal-time-spacing and unequal-time-spacing strategies, respectively. We provides their results in the inset of Fig.~\ref{UnequalTime}. Clearly, the enhancement of the polarization degree is accompanied by the reduction of the spin bath entropy. Also $S$ can be used to demonstrate the power of the unequal-time-spacing strategy. In particular, for the equal-time-spacing strategy, when $N=9$, $S=3.39$, and when $N=20$, $S=2.58$. In sharp contrast, for $L=1$, when $N=9$, $S=0.05$, and when $N=20$, $S\approx10^{-5}$.

\subsection{Polarization performance under the far off-resonant condition}

In this section, the application of the unequal-time-spacing strategy is extended to the far-off-resonant condition. According to Eq.~(\ref{alpham2}), the analytical expression for either $\tau_{\rm opt}$ in Eq.~(\ref{opttauP}) or $\tau_{\rm opt}(t)$ in Eq.~(\ref{unequaltau}) becomes invalid in the presence of a significant $\Delta/\omega_0$. In this case, especially for a typical QD system (see Table~\ref{parameters}), $\tau_{\rm opt}(t)$ can be obtained with a numerical simulation.

\begin{table}[htbp]
\centering
\begin{ruledtabular}
\begin{tabular}{cccccc}
&M ($\times10^2$)&\multicolumn{1}{c}{\textrm{$B$ (G)}}&
\multicolumn{1}{c}{\textrm{$\omega_0$ (MHz)}}&
\multicolumn{1}{c}{\textrm{$\Delta/{\omega_0}$}}&
\multicolumn{1}{c}{\textrm{$g/{\omega_0}$}}\\
\hline
NV$^{(1)}$& 5 & 1000 & 120 & 0.1 & 0.03 \\
NV$^{(2)}$& 5 & 900 & 400 & 0.95 & 0.03 \\
\hline \hline
&M($\times10^3$)&\multicolumn{1}{c}{\textrm{$B$ (G)}}&
\multicolumn{1}{c}{\textrm{$\omega_0$ (GHz)}}&
\multicolumn{1}{c}{\textrm{$\Delta/{\omega_0}$}}&
\multicolumn{1}{c}{\textrm{$g/{\omega_0}$}}\\
\hline
QD$^{(1)}$& 2 & 379 & 5 & 0.999 & 0.016\\
QD$^{(2)}$& 2 & 758 & 10 & 0.999 & 0.008
\end{tabular}
\end{ruledtabular}
\caption{Experimental parameters, including bath size, magnetic-field strength, central-spin frequency, detuning, and coupling strength between the central spin and bath spins, for various NV-center systems~\cite{Sangtawesin2016} and QD systems~\cite{liu2007control,yao2006theory}. For the latter, the gyromagnetic ratio of the electron spin is three orders of magnitude greater than that of the surrounding nuclear spins. Generally, it gives rise to a far-off-resonant condition.}
\label{parameters}
\end{table}

\begin{figure}[htbp]
\centering
\includegraphics[width=0.95\linewidth]{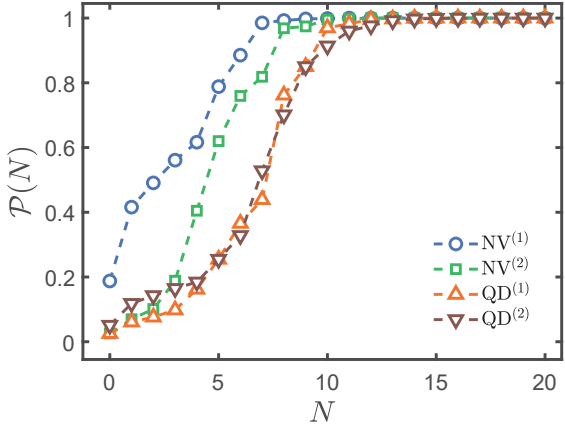}
\caption{Polarization degree of bath spins as a function of measurement number $N$ for NV center systems (blue dashed line with circles and green dashed line with squares) and QD systems (orange dashed line with triangles and the brown dashed line with inverted triangles). $T=0.5$ K, and the other parameters are given by Table~\ref{parameters}.}\label{Bperformance}
\end{figure}

In Fig.~\ref{Bperformance}, we present the performance of the unequal-time-spacing strategy for the four cases listed in Table~\ref{parameters}. For the NV-center systems, we consider both a near-resonant case (see the blue dashed line with circles for NV$^{(1)}$) and a far-off-resonant case (see the green dashed line with squares for NV$^{(2)}$). The initial thermal-state polarization degree for NV$^{(1)}$ is significantly larger than that for NV$^{(2)}$. So the off-resonant case requires more measurements to achieve the same polarization degree as the resonant case. In particular, $\mathcal{P}(8)=0.997$ for NV$^{(1)}$, and $\mathcal{P}(8)=0.968$ for NV$^{(2)}$. While both of them saturate to nearly unit when $N=12$. With more nuclear spins in the bath and the three-order distance in the magnitude of the gyromagnetic ratios for the central and bath spins, the polarization degrees in the QD systems (see the orange dashed line with triangles and the brown dashed line with inverted triangles in Fig.~\ref{Bperformance}) are remarkably lower than those in NV-center systems in the first several rounds. But when $N\geq10$, they can be enhanced to more than $0.91$. In particular, $\mathcal{P}(10)=0.970$ for QD$^{(1)}$, and $\mathcal{P}(10)=0.914$ for QD$^{(2)}$. Moreover, they can be almost completely polarized by $N=15$ measurements.

In the NV-center systems, a nearly complete polarization of nuclear spins was realized by constructing a near-resonant condition around level anti-crossing in the ground state~\cite{Sangtawesin2016,wang2015strongly}, which demands precise control over the external magnetic field ($B\sim 0.1$ T). In the QD systems~\cite{Petersen2013large}, a nearly $50\%$ degree of polarization for nuclear spins could be achieved under a cryogenic temperature ($T\sim 100$ mK) and a strong external magnetic field ($B\sim 2.9$ T). In comparison to the conventional methods, our unequal-time-spacing strategy could achieve complete polarization of nuclear spins with fewer than a dozen rounds of evolution and measurement, in the absence of a precisely controlled external magnetic field or strict ambient conditions, for both NV-center and QD systems.

\section{Discussion}\label{discussion}

\subsection{Success probability}\label{Successprobability}

\begin{figure}[htbp]
\centering
\includegraphics[width=0.95\linewidth]{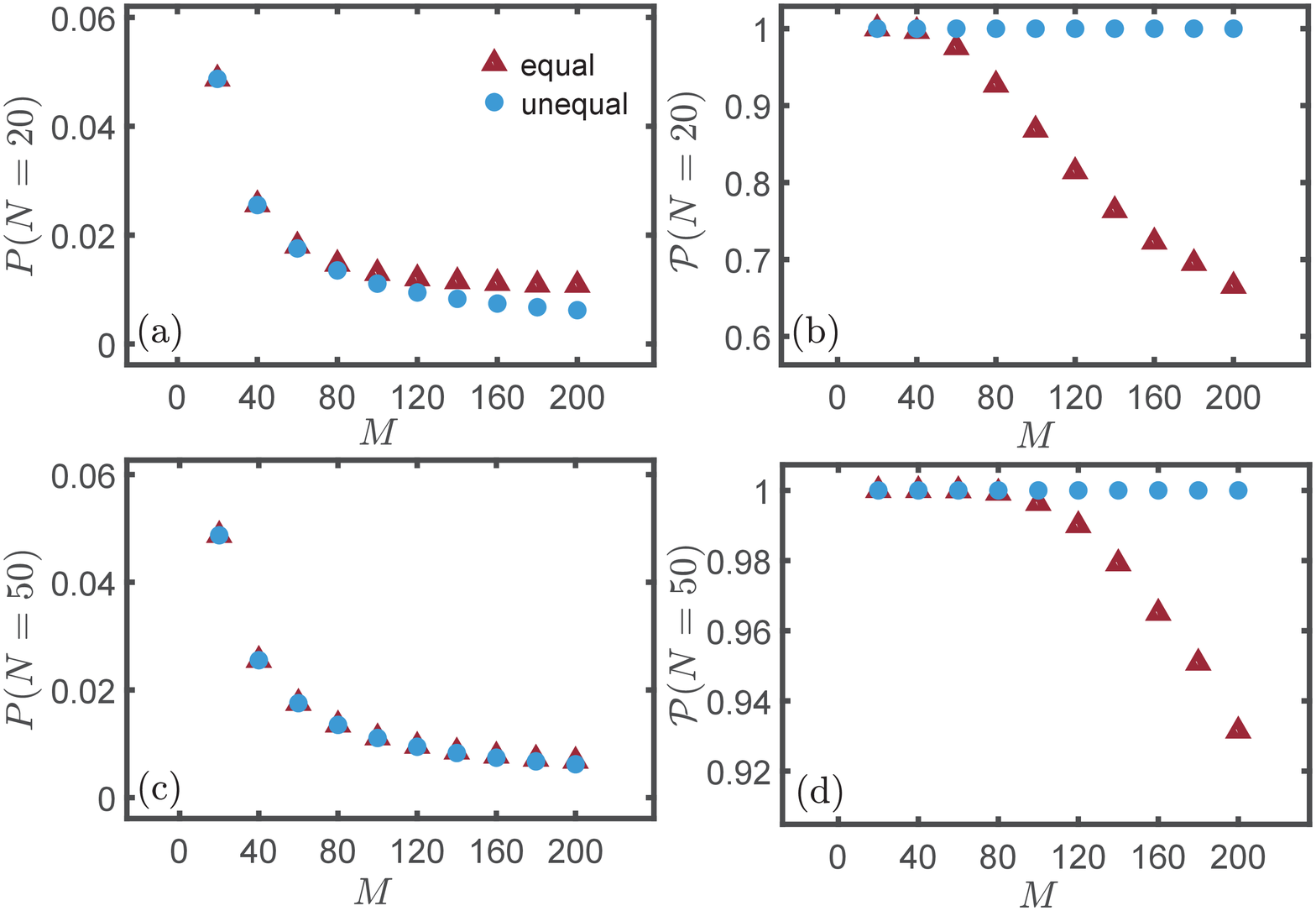}
\caption{(a) and (c) Success probability as a function of the bath size $M$ for $N=20$ and $N=50$ rounds of measurements, respectively. (b) and (d) Polarization degrees corresponding to (a) and (c), respectively. The red triangles and blue circles  represent, respectively, the equal-time-spacing and unequal-time-spacing strategies. The other parameters are the same as in Fig.~\ref{performance}.}\label{comparison}
\end{figure}

The experimental cost for our measurement-induced nuclear spin polarization is described by the success probability in Eq.~(\ref{rhoNtauwithalpha}) or Eq.~(\ref{unequalPg}), since any protocol based on measurement is nondeterministic. In Fig.~\ref{comparison}, we plot the success probabilities and the corresponding polarization degrees for various sizes of bath spins $M$ under both equal-time-spacing and unequal-time-spacing strategies.

We can see that for a lower number of measurements, $N=20$ [Fig.~\ref{comparison}(a)], the success probability of the equal-time-spacing strategy is slightly larger than that of the unequal strategy; however, for a larger number, $N=50$ [Fig.~\ref{comparison}(c)], it is almost invariant for both strategies. Thus, the success probability is insensitive to the optimized measurement interval given by Eqs.~(\ref{opttauP}) and (\ref{unequaltau}). It decreases with $M$ and approaches an asymptotical value of about $1\%$ when $M>160$. In addition, the polarization degree under the unequal-time-spacing strategy is always close to unit, showing advantages over the equal-time-spacing strategy when $M>60$ and $M>100$, as demonstrated in Figs.~\ref{comparison}(b) and ~\ref{comparison}(d), respectively. For the equal-time-spacing strategy, a lower number of measurement, $N=20$, is not enough to polarize a sufficiently large number of nuclear spins. When $M=200$, the success probability $P(N=20)\approx 1.1\%$ with the polarization degree $\mathcal{P}(N=20)=0.67$, and $P(N=50)\approx 0.7\%$ with $\mathcal{P}(N=50)=0.93$.

\subsection{Nonideal interactions between the central spin and nuclear spins}\label{XYandXYZ}

The preceding polarization-by-measurement protocols in our spin-star model are based on the Heisenberg $XY$ interaction, through which they can faithfully exchange the polarized states of the central spin and nuclear spins. In this section, we discuss the effects of two extra interactions, which might present in practical situations, on the polarization performance. The results are obtained under the unequal-time-spacing strategy with the iteratively optimized measurement intervals in Eq.~(\ref{unequaltau}).

\begin{figure}[htbp]
\centering
\includegraphics[width=0.95\linewidth]{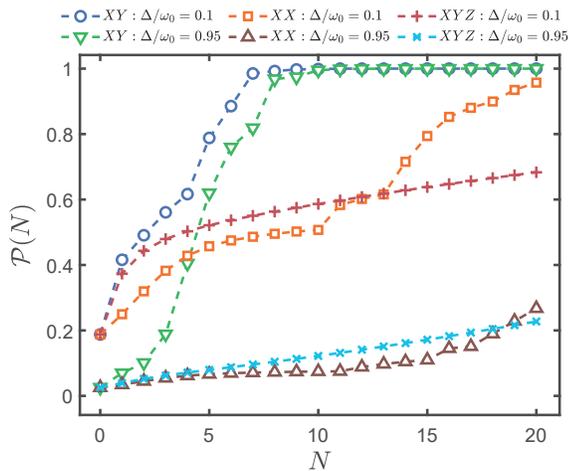}
\caption{Polarization degree of $M=500$ bath spins as a function of the measurement number $N$ in the presence of various interactions and detunings $\Delta/\omega_0$ between the central spin and bath spins. For $XY$ interaction, the blue dashed line with circles and the green dashed line with inverted triangles represent the near-resonant ($\Delta/\omega_0=0.1$) and far-off-resonant cases ($\Delta/\omega_0=0.95$), respectively. For $XX$ interaction, the orange dashed line with squares and the brown dashed line with triangles represent the near-resonant ($\Delta/\omega_0=0.1$) and far-off-resonant cases ($\Delta/\omega_0=0.95$), respectively. For $XYZ$ interaction, the red dashed line with pluses and the cyan dashed line with crosses represent the near-resonant ($\Delta/\omega_0=0.1$) and far-off-resonant cases ($\Delta/\omega_0=0.95$), respectively. $g/\omega_0=0.03$, $T=0.5$ K, and $\omega_0=120$ MHz. }\label{counterrotating}
\end{figure}

First, we consider the Heisenberg $XX$ interaction between the central spin and bath spins, which is equivalent to including the high-frequency-modulated counter-rotating terms into the flip-flop interaction in the interaction picture. Then using the collective angular momentum operators, the full Hamiltonian in the Schr\"odinger picture can be written as
\begin{equation}\label{XX Ham1}
\begin{aligned}
H&=H_0+H_I,\\
H_0&=\frac{\omega_0}{2}\sigma_d^z+\omega_1J_z,\\
H_I&=2g\left(\sigma_d^+J_-+\sigma_d^-J_+\right)+2g\left(\sigma_d^+J_++\sigma_d^-J_-\right).
\end{aligned}
\end{equation}
The last counter-rotating terms in $H_I$ are conventionally neglected when $g\ll\omega_0, \omega_1, |\Delta|$.

Second, we can consider the Heisenberg $XYZ$ interaction; that is, there is a longitudinal interaction in addition to the transverse interaction between the central spin and the bath spins. In the rotating frame with respect to $H'_0=\omega_1/2(\sigma_d^z+J_z)$, the full Hamiltonian can be written as
\begin{equation}\label{IntHamXYZ}
H'=\frac{\Delta}{2}\sigma_d^z+2g\left(J_+\sigma_d^-+J_-\sigma_d^+\right)+gJ_z\sigma_d^z,
\end{equation}
where the longitudinal interaction strength is set to be the same as the transverse one for simplicity.

In Fig.~\ref{counterrotating}, we demonstrate the polarization performances for various types of interactions and detunings $\Delta/\omega_0$ within $N=20$ rounds of measurements. With a fixed number of bath spins, a smaller detuning gives rise to a larger thermal-state polarization $\mathcal{P}_{\rm th}$ and also better polarization performance for any interaction between the central spin and bath spins. The presence of either counterrotating interaction or longitudinal interaction always suppresses the polarization effect by our measurement protocol, which becomes dramatically severe in the far-off-resonant situation. In particular, in the near-resonant case $\Delta/\omega_0=0.1$, it is found that $\mathcal{P}(8)=0.99$ for $XY$ interaction, $\mathcal{P}(20)=0.96$ for $XX$ interaction, and $\mathcal{P}(20)=0.68$ for $XYZ$ interaction. In sharp contrast, when $\Delta/\omega_0=0.95$, $\mathcal{P}(10)=0.99$ for $XY$ interaction, $\mathcal{P}(20)=0.27$ for $XX$ interaction, and $\mathcal{P}(20)=0.23$ for $XYZ$ interaction. However, in a weak-coupling regime $g/\omega_0=0.03$, the presence of the counter-rotating interaction can not be ignored, especially under a far-off-resonant condition. When $N\leq13$, the polarization degree of the bath spins under $XYZ$ interaction is higher than that under $XX$ interaction. Roughly, the suppression effect from the longitudinal interaction is more severe than that from the counter-rotating interaction.\\

\section{Conclusion}\label{Conclusion}

In summary, we proposed a measurement-based dynamical nuclear-spin polarization protocol in a spin-star model, where the central spin is coupled to the surrounding bath spins with the Heisenberg $XY$ interaction. The mean-field approach and the permutational invariance of the bath spins allow us to use collective angular momentum operators to model the behavior of our model, similar to the semianalytical simulation over the spin-spin-environment configuration. The central spin and the bath spins were prepared in the ground state and the thermal equilibrium state, respectively. A nearly $100\%$ polarization of the bath spins was realized by repeated instantaneous projective measurements performed on the ground state of the central spin. The key idea is that the ground states of the central spin and bath spins are closely connected under the interaction Hamiltonian. The polarization performance can be dramatically increased by iteratively optimizing the measurement interval $\tau_{\rm opt}$, which is determined by the polarization degree at the end of the last round of evolution and measurement, the size of the spin bath, and the coupling strength between the central spin and bath spins. As the cost of our nondeterministic protocol, the success probability is found to be insensitive to the measurement interval.

Our protocol applies to both near-resonant and far-off-resonant conditions between central spin and bath spins. With an ideal unequal-time-spacing strategy, the nuclear spins could be completely polarized in fewer than 20 measurements for both NV-center and QD systems. They are scalable solid-state systems and good candidates for various quantum technology applications.

Our polarization-by-measurement protocol can be considered an extension of measurement-based cooling that originates from cooling mechanical oscillators in optomechanics, in which repeated measurements on the ground state of the ancillary system generate fast cooling with a finite success probability. Our method is different from the quantum Zeno effect, for which it was predicted and verified that frequent and controlled measurements into a fixed state or subspace can inhibit a quantum system from leaving that state or subspace. Rather than taking the measurement interval to zero in the limit for the quantum Zeno effect, we find that the probability in which a scalable and randomly aligned spin system stays at the polarized state or subspace can be stably accumulated through measurements with optimized intervals.

\begin{figure}[htbp]
\centering
\includegraphics[width=0.95\linewidth]{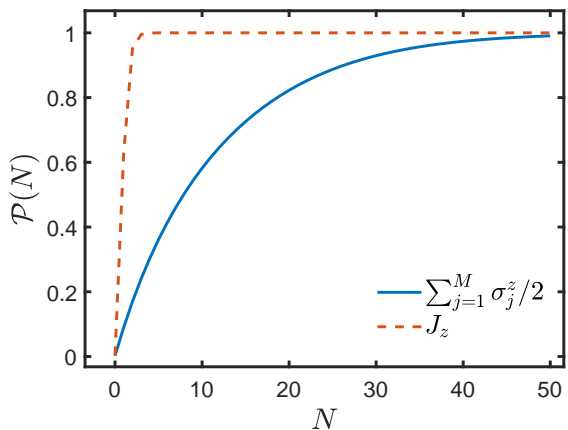}
\caption{Polarization degree of bath spins $\mathcal{P}(N)$ as a function of the number of measurements $N$ for the unequal-time-spacing strategy. The blue solid line and the red dashed line represent the results for treating the bath spins as individual spins and a collective large spin, respectively. The other parameters are the same as in Fig.~\ref{performance}.}\label{comparisonJzSz}
\end{figure}

\section*{Acknowledgments}

We acknowledge financial support from the National Science Foundation of China (Grants No. 11974311 and No. U1801661).

\appendix
\section{Comparing large spin and individual spins}\label{effectivecoarse}

To justify the applicability of our polarization-by-measurement protocol in the subspace with $J=M/2$, here we present numerically the polarization performances for a model with $M=8$ bath spins. As shown in Fig.~\ref{comparisonJzSz}, they are exactly calculated in the whole Hilbert space and the $J=M/2$ subspace. It is shown that the bath spins can still be fully polarized while being treated as individual spins. This requires more rounds of measurements in comparison to those needed when treating the bath spins as a collective large spin. In particular, the latter treatment requires $N<10$, and the former requires about $N\approx50$ to achieve $\mathcal{P}\approx0.99$. The weight of the coarse-grained subspaces increases surely with the size of bath spins, which costs more resources. Yet it is not crucial to the proof of principle for our polarization protocol.

\bibliographystyle{apsrevlong}

\bibliography{ref}

\end{document}